\documentclass[english,aps,preprint]{revtex4}
\usepackage[T1]{fontenc}
\usepackage[latin9]{inputenc}
\usepackage{array}
\usepackage{float}
\usepackage{multirow}
\usepackage{amsmath}
\usepackage{amssymb}
\usepackage{graphicx}

\makeatletter

\providecommand{\tabularnewline}{\\}

\@ifundefined{textcolor}{}
{%
 \definecolor{BLACK}{gray}{0}
 \definecolor{WHITE}{gray}{1}
 \definecolor{RED}{rgb}{1,0,0}
 \definecolor{GREEN}{rgb}{0,1,0}
 \definecolor{BLUE}{rgb}{0,0,1}
 \definecolor{CYAN}{cmyk}{1,0,0,0}
 \definecolor{MAGENTA}{cmyk}{0,1,0,0}
 \definecolor{YELLOW}{cmyk}{0,0,1,0}
}

\makeatother

\usepackage{babel}
\begin{document}

\title{Search for excited muons at the future SPPC-based muon-proton colliders}

\author{A. Caliskan}
\email{acaliskan@gumushane.edu.tr}

\affiliation{Gümü\c{s}hane University, Faculty of Engineering and Natural Sciences,
Department of Physics Engineering, 29100, Gümü\c{s}hane, Turkey}
\begin{abstract}
We have investigated the production potential of the spin-1/2 excited
muons, predicted by the preonic models, at the four SPPC (Super Proton-Proton
Collider)-based muon-proton colliders in different center-of-mass
energies. For the signal process $\mu p\rightarrow\mu^{\star}X\rightarrow\mu\gamma X$,
the production cross-section and the decay width values of the excited
muons have been calculated. The pseudorapidity and transverse momentum
distributions of the final state particles of muons and photons have
been obtained to be determine the kinematical cuts best suited for
discovery of the excited muons. By applying these cuts we have gotten
the discovery, observation and exclusion mass limits of the excited
muons for two compositeness scale values. It is shown that the discovery
limits on the excited muons in case the compositeness scale equals
to $100$ TeV are $2.7$, $3.9$, $3.1$ and $6.7$ TeV for center-of-mass
enerjies of $10.3$, $14.2$, $14.6$ and $20.2$ TeV, respectively. 
\end{abstract}
\maketitle

\section{\i ntroduction}

In the particle physics all matter and antimatter consist of two kinds
of elementary particles: leptons and quarks. The electromagnetic,
weak and strong forces among these particles are described very well
by the Standard Model (SM). Despite all this success, there are some
issues which have not entirely solved by the SM, such as large number
of free parameters, fermion mass generation and fermion mixing, CP
violation and the electroweak symmetry breaking mechanism. A lot of
alternative theories beyond the SM (BSM), like Technicolour \cite{weinberg,susskind},
Grand Unified Models \cite{unity of all,pati-salam}, Supersymmetry
\cite{extension of the}, Compositeness \cite{I.A. DSouza}, have
been proposed to solve these issues. 

One of the most important BSM models is the compositeness that explains
the inflation of elementary particles and quark-lepton symmetry in
the best manner, introducing the more fundamental constituents in
which the leptons, quarks and their antiparticles have a substructure
called preons. Countless preonic models have been suggested so far,
by the particle physicists, and some new types of particles are introduced
in the framework of these models, such as excited fermions, leptoquarks,
leptogluons and color sextet quarks. The simplest preon model, for
example, is Harari-Shupe model \cite{Harari,Shupe} in which leptons
and quarks are bound states of two elementary, massles spin-1/2 preons,
called T- and V-rishons, which cary both color and hypercolor charges.
New hypercolor interactions among the fermions should exist at the
scale of energies that bind the preons together. This energy scale
is an important parameter that is commonly used in the all composite
models, and called the compositeness scale, $\Lambda$. The quarks
and leptons are composed of three preons for this model, and they
are hypercolor singlets below the $\Lambda$ while the all preons
become confined.

If the SM fermions are composite, they must have the excited states
as a result of the compositeness. Therefore the excited fermions could
be considered as the excited state of the SM fermions that are in
ground state and should be observed experimentally. The excited fermions
are predicted by the preonic models. They were firstly proposed in
1977, and studied in detail in the following years \cite{unified model,subquark model,observable,a fundamental}.
The excited leptons and quarks can have spin-1/2 and spin-3/2 states,
and it is expected that their masses are heavier than ones of the
SM fermions. In the present study we have interested in the excited
muons with spin-1/2 as a continuation of our recent works about the
excited leptons \cite{excited muon searches,excited neutrino search}.
There are also considerable phenomenological studies on the excited
quarks \cite{g=0000FCnayd=000131n} and leptons \cite{search for excited electrons,analysis of excited neutrinos,V.Ar=000131}
in the literature. 

Even though no signals for the excited leptons can be found in the
experimental studies at the LEP \cite{LEP}, HERA \cite{HERA}, TEVATRON
\cite{Tevatron}, ATLAS \cite{ATLAS} and CMS \cite{CMS} experiments,
the more powerful accelerators that will be established in the future
will continue to be hopeful for their discovery. A possible discovery
of the excited leptons will provide a direct evidence of the excited
lepton compositeness. The recent experimental mass limits on the excited
muons for pair and single production are provided by the OPAL and
ATLAS collaborations, respectively \cite{particle data group}. Exclusion
limits of the excited muons are $m_{\mu^{\star}}>103.2$ GeV for $e^{+}e^{-}\rightarrow\mu^{\star}\mu^{\star}$
and $m_{\mu^{\star}}>3000$ GeV for $pp\rightarrow\mu\mu^{\star}X$
, assuming the coupling parameter of $f=f^{\prime}=1$ and the energy
scale of $\varLambda=m_{\mu^{\star}}$. The excited muons can be also
obtained via contact interactions, but in this paper we only concentrated
the gauge interactions. The four fermion contact interactions have
a different Lagrangian in itself, and it may be studied in the framework
of the another study.

We have investigated the excited muon production at the four different
center-of-mass energies of the SPPC-based muon-proton colliders. We
present the SPPC-based muon-proton colliders and their main parameters
in the Section 2, the excited muon interaction Lagrangian, its decay
widths and the cross-sections in the Section 3, and the signal-background
analysis in the Section 4. Finally, we have reported the all results
in the last section.

\section{THE SPPC-BASED MUON-PROTON COLLIDERS}

The discovery of the Higgs particle with a mass of 125 GeV at the
Large Hadron Colider (LHC) in 2012 \cite{ATLAS-ke=00015Fif,CMS-ke=00015Fif}
confirmed the electroweak symmetry breaking mechanism of the SM. With
this discovery the particle physics has reached the Higgs era, but
it is not known whether the observed Higgs particle is the fundamental
scalar. To study the properties of the Higgs boson in detail and understand
its true nature, the world high-energy physics community has started
to investigate the feasibility of a Higgs factory. It is known that
the hadron colliders like the LHC have the highest center-of-mass
energy values, so they are called the discovery machines, while the
lepton and lepton-hadron colliders provide the smaller ones, and called
the precision machines. Recently, various future collider projects
have begun to be designed by the accelerator physicists such as the
ILC (International Linear Collider) \cite{ILC}, LHeC (Large Hadron
Electron Collider) \cite{LHeC web} and CLIC (Compact Linear Collider)
\cite{CLIC} to search primarily the Higgs physics. 

In the post-LHC era the most important collider project in Europe
is no doubt the international Future Circular Collider (FCC) \cite{FCC web}
with a center-of-mass energy of 100 TeV, has been launched in 2010-2013
at the CERN, and supported by the European Union within the Horizon
2020 Frramework for Research and Innovation. The main purpose of this
project is to establish a 100 TeV energy-frontier hadron collider
(FCC-hh) to be allocated in a new 80-100 km tunnel at the CERN. As
an intermediate step of the FCC project, it also involves a high-luminosity
lepton collider (FCC-ee or TLEP \cite{TLEP project web page}) with
a center-of mass energy of 90-400 GeV, to be installed in the same
tunnel, as well as a lepton-hadron collider option (FCC-he). The FCC
will give us the opportunity to explore the properties of Higgs boson,
new interactions beyond the SM, top quark et cetera, at the highest
energies. The CDR (Conceptual Design Report) of the FCC is expected
to be completed in 2018.

In parallel with the developments related to the FCC project in Europe,
the chinese physicist have initiated the design study of a two-stage
circular collider project in 2012, called the CEPC-SPPC. The first
stage of the project is a circular electron-positron collider (CEPC)
with a center-of-mass energy of 240 GeV, to search the properties
of Higgs particle. After completing its missions, the CEPC will be
upgraded to the second stage that is a Super Proton-Proton Collider
(SPPC) with a center-of-mass energy of more than 70 TeV, aiming at
researching the BSM physics. The Preliminary Conceptual Design Report
(Pre-CDR) of the CEPC-SPPC project has been completed by the CEPC-SPPC
study group in 2015 \cite{CEPC-CDR}. By using the same tunnel as
the CEPC that is 54.7 km in circumference, center-of-mass energy of
about 70 TeV will be tried to be reached. But, larger circumference
options for the SPPC collider are also being considered. Table 1 presents
the main parameters for the all design options of the SPPC collider
\cite{SPPC Parameter}.

\begin{table}
\caption{The main parameters of proton beams in the SPPC collider for the various
design options.}

\begin{tabular}{|c|c|c|c|c|c|}
\hline 
Parameters & Option-1 (Pre-CDR) & Option-2 & Option-3 & Option-4 & Option-5\tabularnewline
\hline 
\hline 
Beam energy (TeV) & $35.6$ & $35$ & $50$ & $68$ & $50$\tabularnewline
\hline 
Circumference (km) & $54.7$ & $54.7$ & $100$ & $100$ & $78$\tabularnewline
\hline 
Dipole field (T) & $20$ & $19.69$ & $14.73$ & $20.03$ & $19.49$\tabularnewline
\hline 
Peak luminosity ($x10^{35}cm^{-2}s^{-1}$) & $1.1$ & $1.2$ & $1.52$ & $10.2$ & $1.52$\tabularnewline
\hline 
Particle per bunch ($10^{11})$ & $2$ & $2$ & $2$ & $2$ & $2$\tabularnewline
\hline 
Norm. transverse emittance ($\mu m$)  & $4.1$ & $3.72$ & $3.65$ & $3.05$ & $3.36$\tabularnewline
\hline 
Bunch number per beam & $5835$ & $5835$ & $10667$ & $10667$ & $8320$\tabularnewline
\hline 
Bunch length (mm) & $75.5$ & $56.5$ & $65$ & $15.8$ & $70.6$\tabularnewline
\hline 
Bunch spacing (ns) & $25$ & $25$ & $25$ & $25$ & $25$\tabularnewline
\hline 
\end{tabular}
\end{table}

If a TeV energy muon collider is installed tangentially to the SPPC,
a muon-proton collider, at the high center-of-mass energy, can be
obtained. Taking into account the energy values of 0.75 and 1.5 TeV
for muon beam, and design options of 35.6 and 68 TeV of the SPPC,
four muon-proton collider options have been recently proposed \cite{SPPC-based}.
The excited muon production potential at the these four muon-proton
colliders have been analyzed in this paper, and its basic parameters
are shown in the Table 2.

\begin{table}

\caption{The main parameters of the SPPC-based muon-proton colliders.}

\begin{tabular}{|c|c|c|c|c|c|c|}
\hline 
Colliders & $E_{\mu}$$\left(TeV\right)$ & $E_{p}$$\left(TeV\right)$ & $\sqrt{s}$ $\left(TeV\right)$ & $L_{int}$ $\left(fb^{-1}\right)$ & $\xi_{\mu}$ & $\xi_{p}$\tabularnewline
\hline 
\hline 
$\mu750$-SPPC1 & $0.75$ & $35.6$ & $10.33$ & $5.5$ & $8.7x10^{-3}$ & $6x10^{-2}$\tabularnewline
\hline 
$\mu750$-SPPC2 & $0.75$ & $68$ & $14.28$ & $12.5$ & $8.7x10^{-3}$ & $8x10^{-2}$\tabularnewline
\hline 
$\mu1500$-SPPC1 & $1.5$ & $35.6$ & $14.61$ & $4.9$ & $8.7x10^{-3}$ & $6x10^{-2}$\tabularnewline
\hline 
$\mu1500$-SPPC2 & $1.5$ & $68$ & $20.2$ & $42.8$ & $8.7x10^{-3}$ & $8x10^{-2}$\tabularnewline
\hline 
\end{tabular}

\end{table}

\section{THE EXCITED MUONS }

The interactions of an excited lepton with ordinary leptons are of
magnetic transition type, and the effective Lagrangian that describes
the interaction between a spin-1/2 excited lepton, the SM lepton and
a gauge boson is given as \cite{excited lepton production}\cite{excited-quark}\cite{looking for}\cite{excited fermions},

\begin{equation}
L=\frac{1}{2\Lambda}\overline{l_{R}^{\star}}\sigma^{\mu\nu}\left[fg\frac{\vec{\tau}}{2}\centerdot\vec{W}_{\mu\nu}+f^{\prime}g^{\prime}\frac{Y}{2}B_{\mu\nu}\right]l_{L}+h.c.,
\end{equation}
where $l$ and $l^{\star}$ represent the SM lepton and the excited
lepton, respectively, $\Lambda$ is the new physics scale, $\overrightarrow{W}_{\mu\nu}$
and $B_{\mu\nu}$ are the field strength tensors, $g$ and $g^{\prime}$
are the SM gauge couplings of SU(2) and U(1), $f$ and $f^{\prime}$
are the new scaling factors for the gauge couplings, Y is hypercharge,
$\sigma^{\mu\nu}=i(\gamma^{\mu}\gamma^{\nu}-\gamma^{\nu}\gamma^{\mu})/2$
where $\gamma^{\mu}$ are the Dirac matrices, and $\overrightarrow{\tau}$
denotes the Pauli matrices.

The excited muons can decay into three channels that are $\gamma$
- channel ($\mu^{\star}\rightarrow\mu\gamma$), Z - channel ($\mu^{\star}\rightarrow\mu Z$)
and W-channel ($\mu^{\star}\rightarrow\mu W$). The decay widths of
the excited muons are given the following formula,

\begin{equation}
\Gamma(l^{\star}\rightarrow lV)=\frac{\alpha m^{\star3}}{4\Lambda^{2}}f_{V}^{2}(1-\frac{m_{V}^{2}}{m^{\star2}})^{2}(1+\frac{m_{V}^{2}}{2m^{\star2}}),
\end{equation}
where $m^{\star}$ is the mass of the excited electron, $m_{V}$ is
the mass of the gauge boson, $f_{V}$ is the new electroweak coupling
parameter corresponding to the gauge boson $V$, where $V=$$W$,
$Z$, $\gamma$, and $f_{\gamma}=-(f+f^{\prime})/2,$ $f_{Z}=(-f\cot\theta_{W}+f\tan\theta_{W})/2,$
$f_{W}=(f/\sqrt{2}\sin\theta_{W})$, where $\theta_{W}$ is the weak
mixing angle, and $\alpha$ is the electromagnetic coupling constant.

\begin{figure}
\begin{centering}
\includegraphics[scale=0.85]{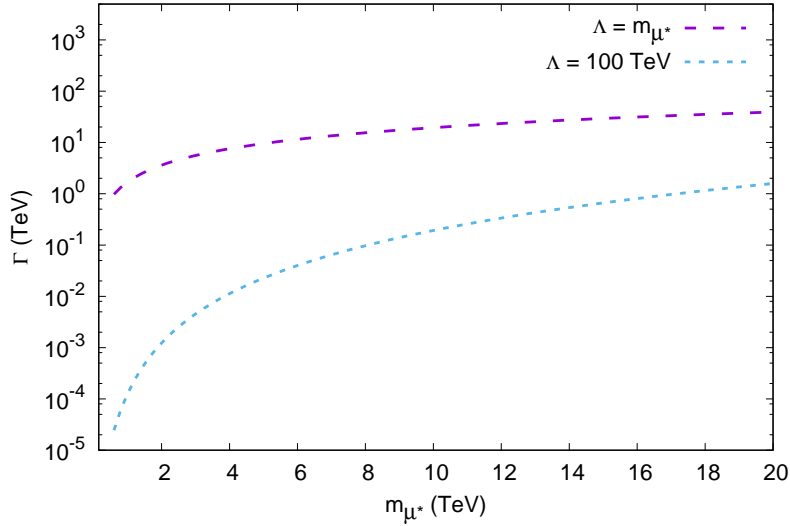}
\par\end{centering}
\caption{The total decay widths of the excited muons for $\varLambda=m_{\mu^{\star}}$
and $\varLambda=100$ TeV, assuming $f=f^{\prime}=1$ .}
\end{figure}

For the numerical calculations we implemented the excited muon interaction
vertices into the high-energy simulation programme of CALCHEP \cite{calchep}.
Figure 1 shows the total decay widths of the excited muons for energy
scales $\varLambda=m_{\mu^{\star}}$ and $\varLambda=100$ TeV. The
total cross-sections of the excited muons produced at the four different
muon-proton colliders, which are $\mu750$-SPPC1, $\mu750$-SPPC2,
$\mu750-SPPC2,$ $\mu1500$-SPPC1, $\mu1500$-SPPC2, are shown in
the Figure 2, using the program CALCHEP with the CTEQ6L parton distribution
functions \cite{PDF}. It is unambiguously seen from the figure that
the excited muons have sufficiently high cross-sections for both energy
scales.

\begin{figure}
\includegraphics[scale=0.6]{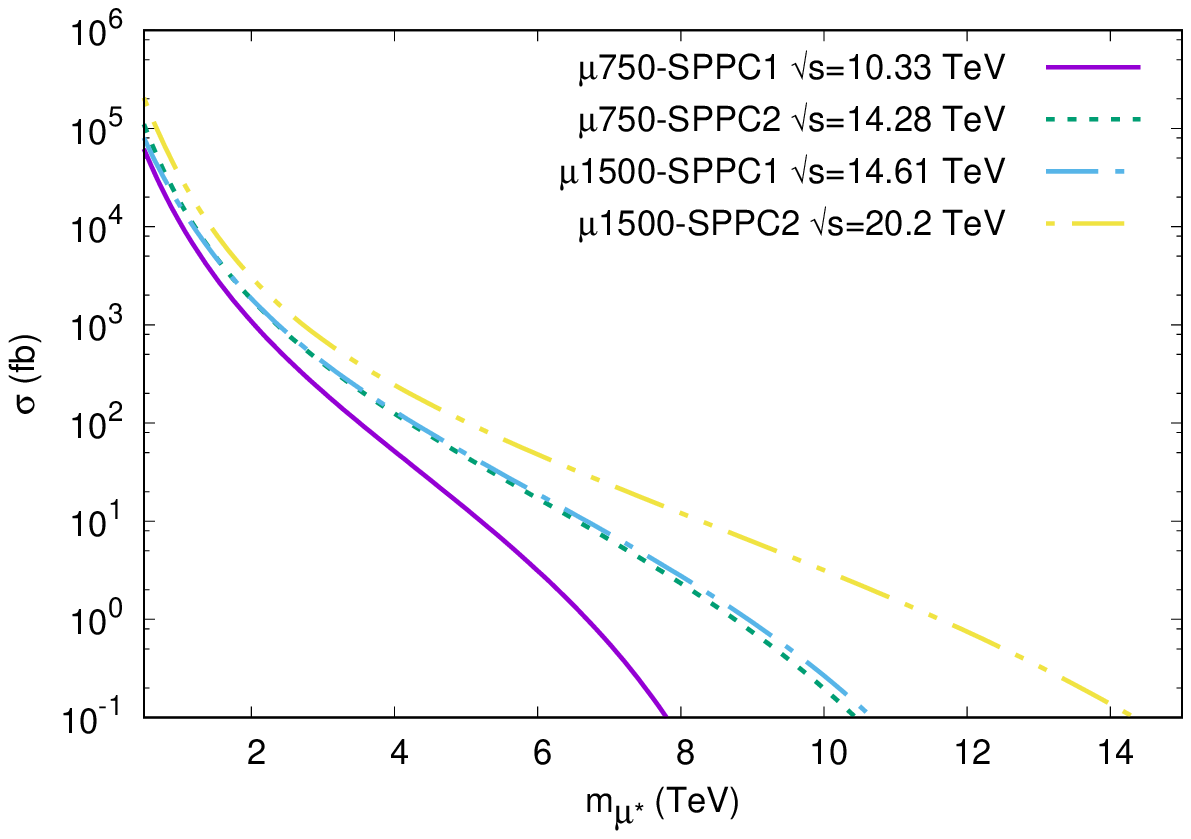}\includegraphics[scale=0.6]{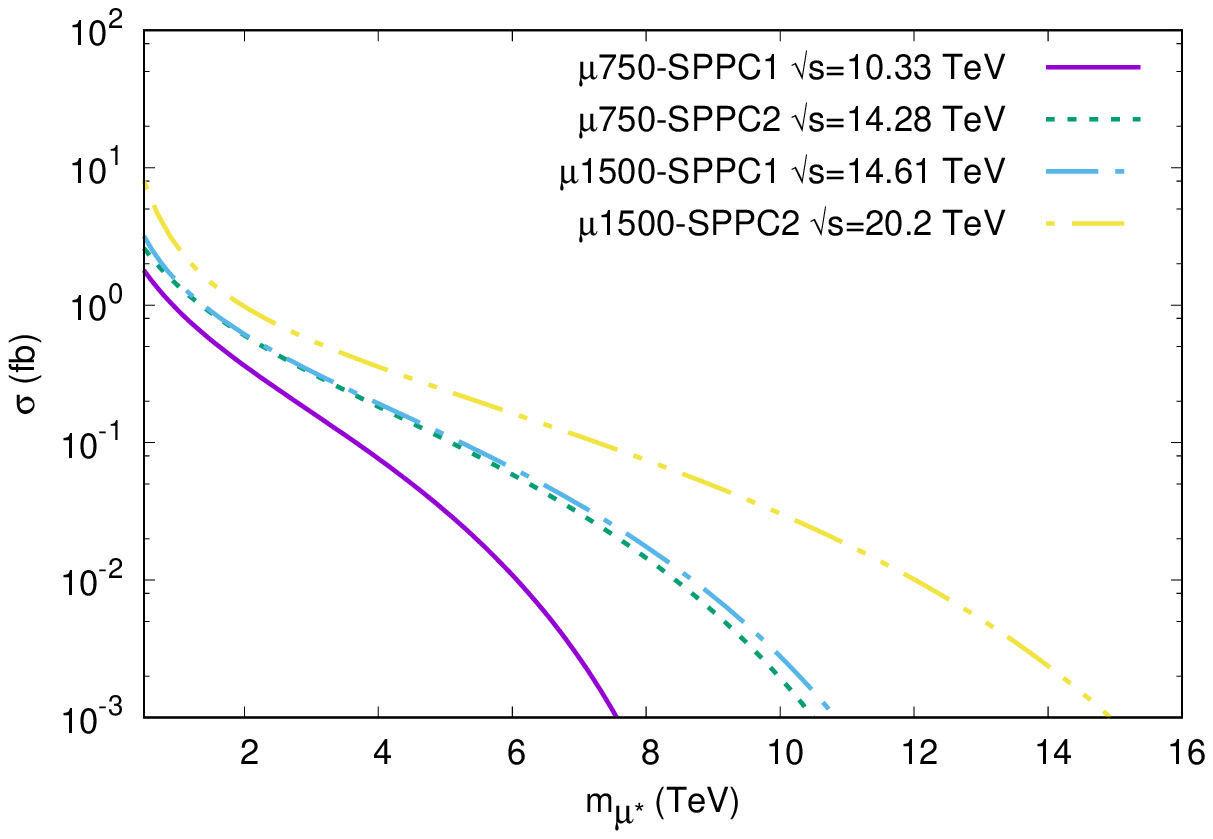}

\caption{The total cross-section values of the excited muons with respect to
its mass at the various muon-proton colliders for $\varLambda=m_{\mu^{\star}}$(left)
and $\varLambda=100$ TeV (right), assuming $f=f^{\prime}=1$.}

\end{figure}

\section{SIGNAL AND BACKGROUND ANALYSIS}

The SPPC-based muon-proton colliders will allow us to explore the
excited muons via $\mu p\rightarrow\mu^{\star}X$ process with subsequent
decays of the excited muons into a muon and photon. Therefore, our
signal process is $\mu p\rightarrow\mu\gamma X$, and subprocesses
are $\mu q(\overline{q})\rightarrow\mu\gamma q(\overline{q})$. In
order to separate the excited muon signals from the background we
have applied prime transverse momentum cuts to the final state particles
of muon, photon and jets, as $p_{T}^{\mu,\gamma,j}>20$ GeV. The SM
cross-section values after the application of these prime cuts have
been obtained as $\sigma_{B}=$73.15 pb for $\mu750$-SPPC1, $\sigma_{B}=$82.11
pb for $\mu750$-SPPC2, $\sigma_{B}=$84.65 pb for $\mu1500$-SPPC1,
$\sigma_{B}=$126.84 pb for $\mu1500$-SPPC2 collider. To be assign
the kinematical cuts best suited for the discovery of the excited
muons, we have to look at the transverse momentum ($p_{T}$) and pseudorapidity
($\eta$) distributions of the final state particles for both the
signal and background.

\begin{figure}
\includegraphics[scale=0.45]{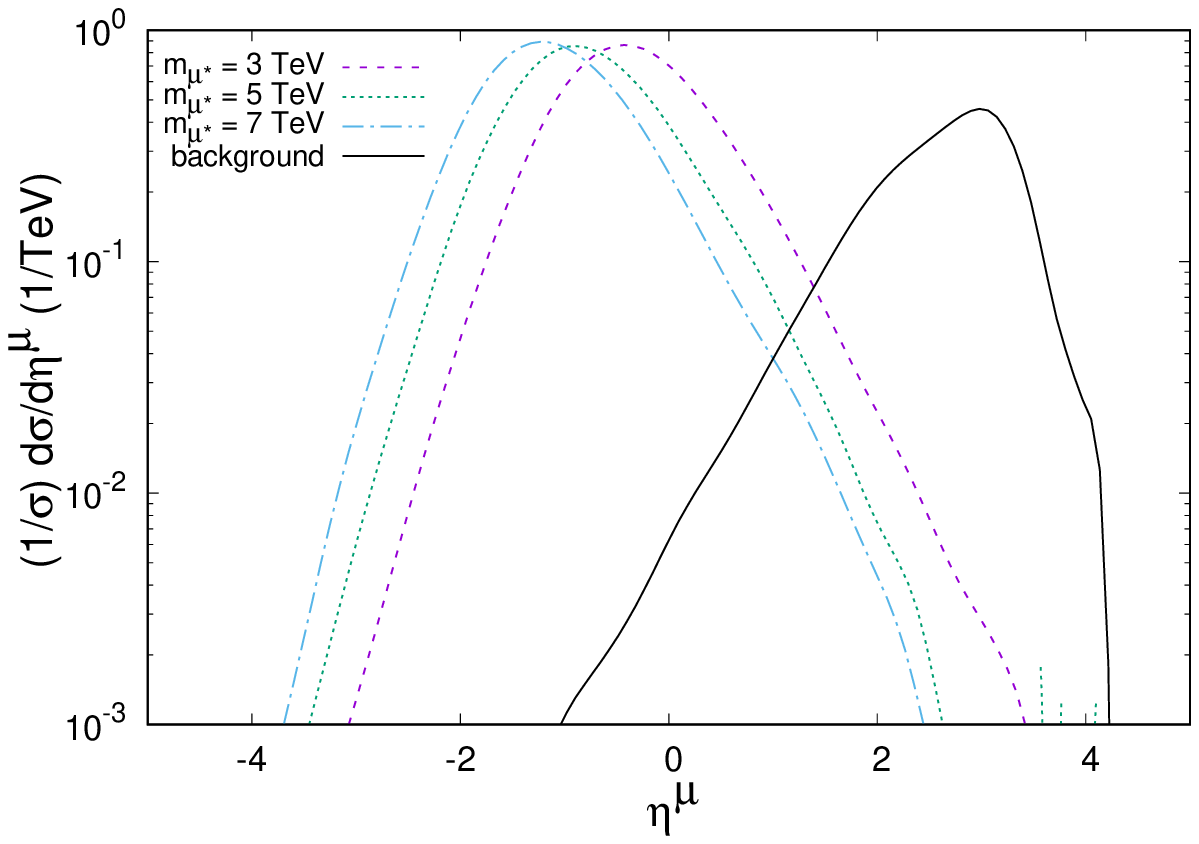}\includegraphics[scale=0.45]{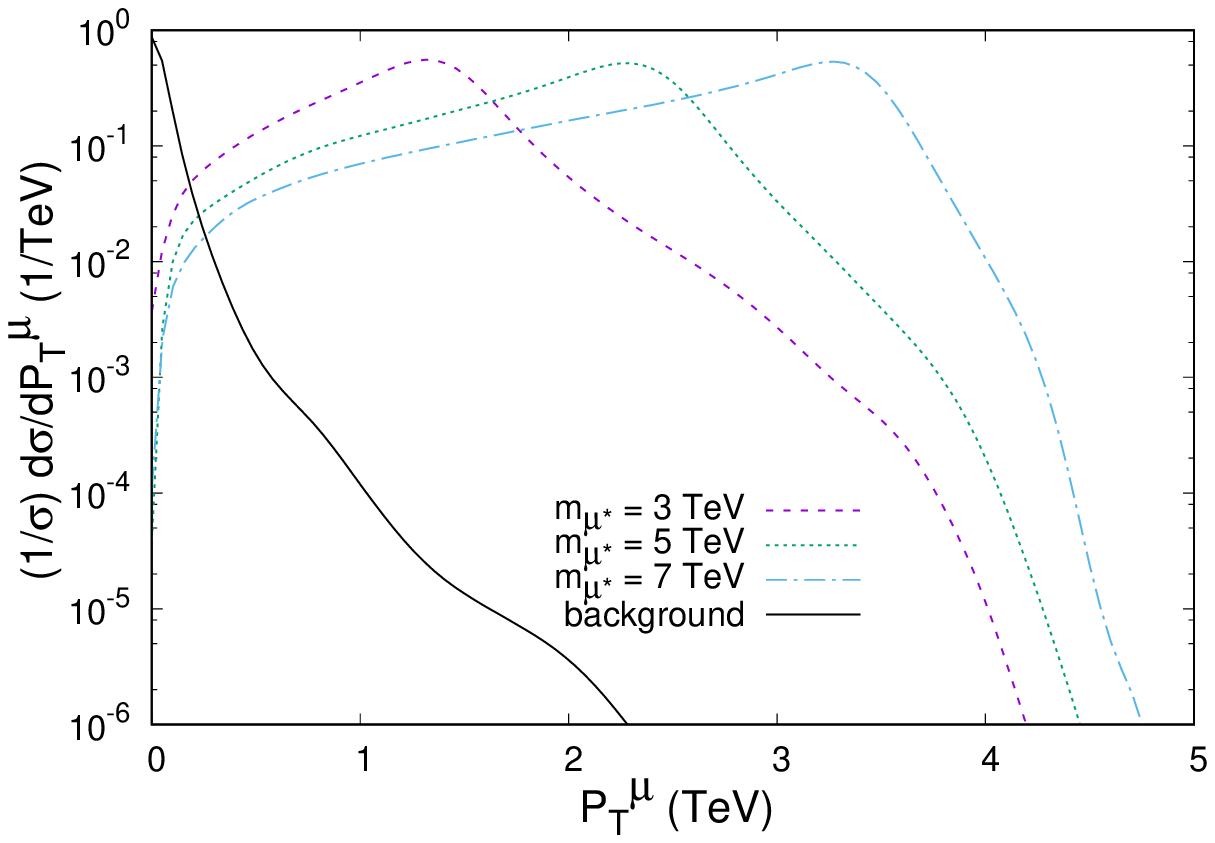}\includegraphics[scale=0.45]{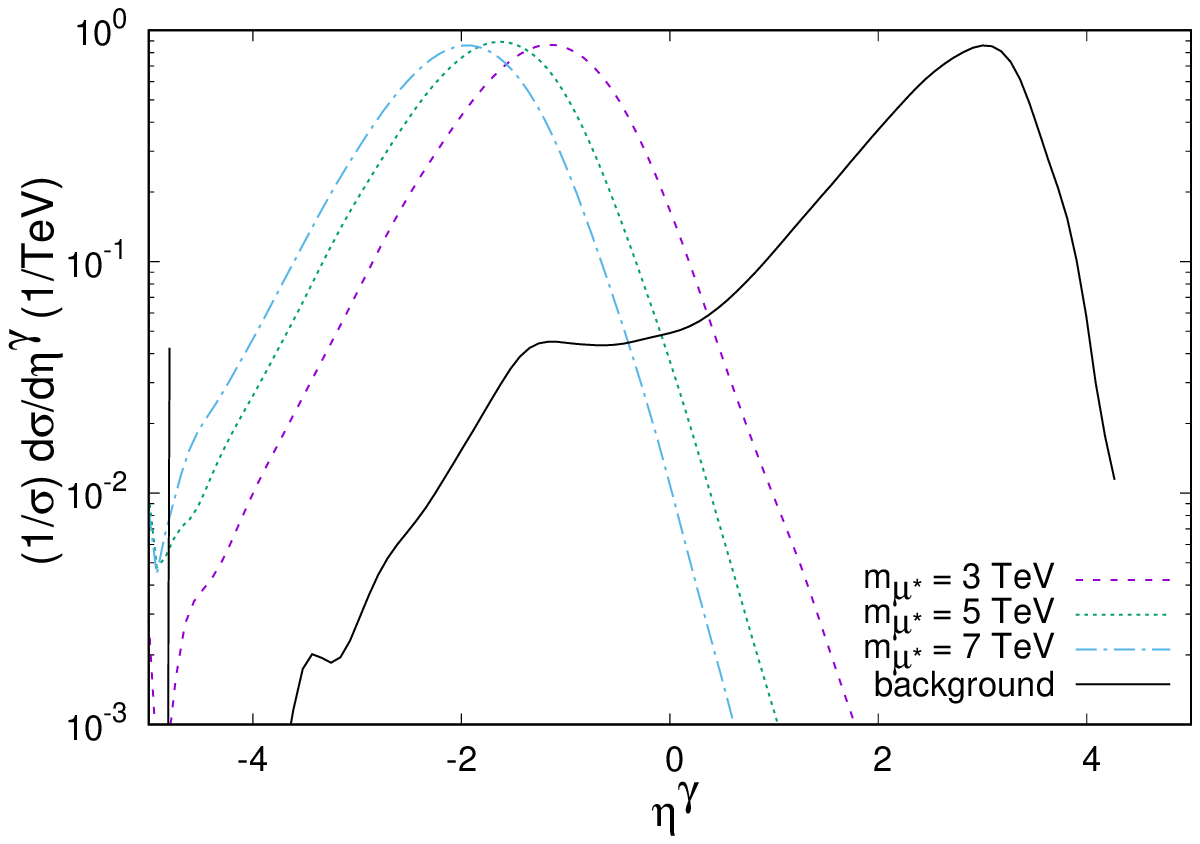}

\caption{The normalized pseudorapidity (left) and transverse momentum (middle)
distributions of the final state muons and the normalized pseudorapidity
distributions of the final state photons (right) at the $\mu750$-SPPC1
collider, for $f=f^{\prime}=1$ and $\varLambda=m_{\mu^{\star}}$.}

\end{figure}

\begin{figure}
\includegraphics[scale=0.45]{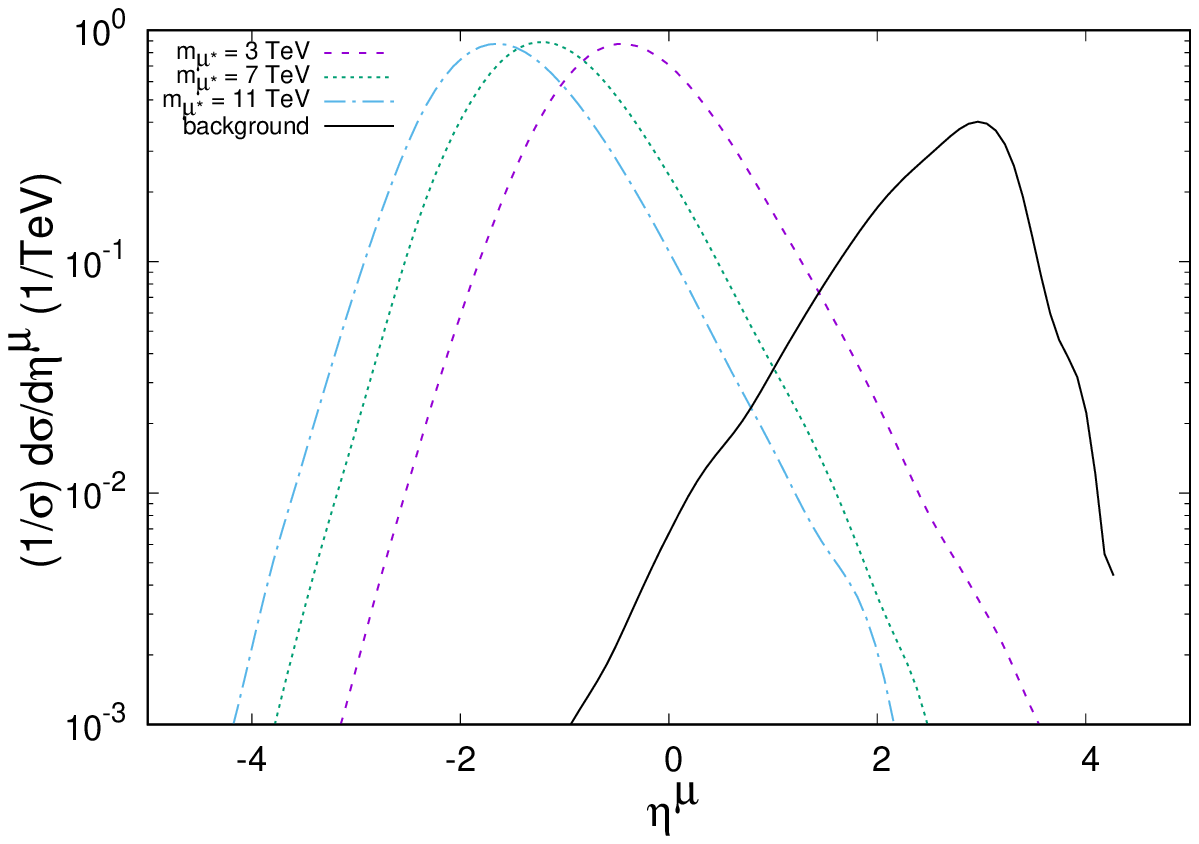}\includegraphics[scale=0.45]{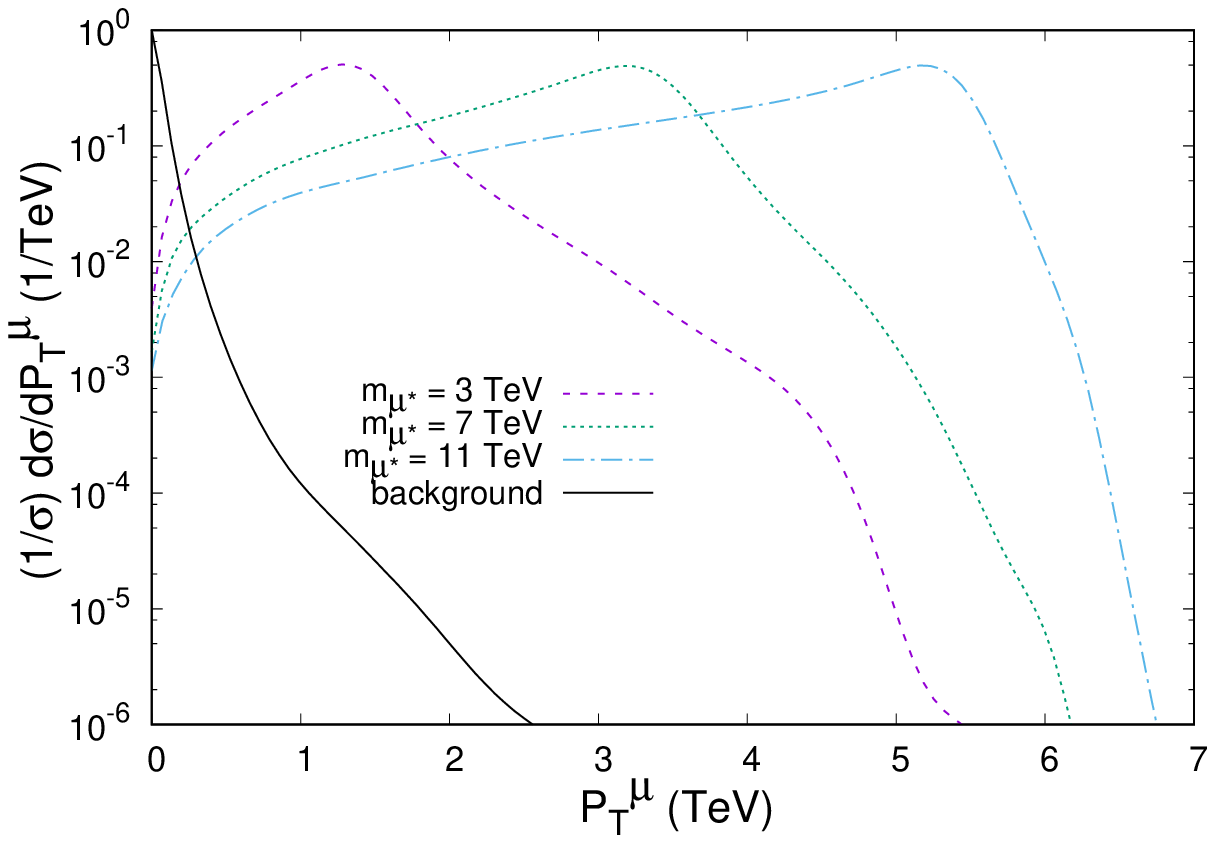}\includegraphics[scale=0.45]{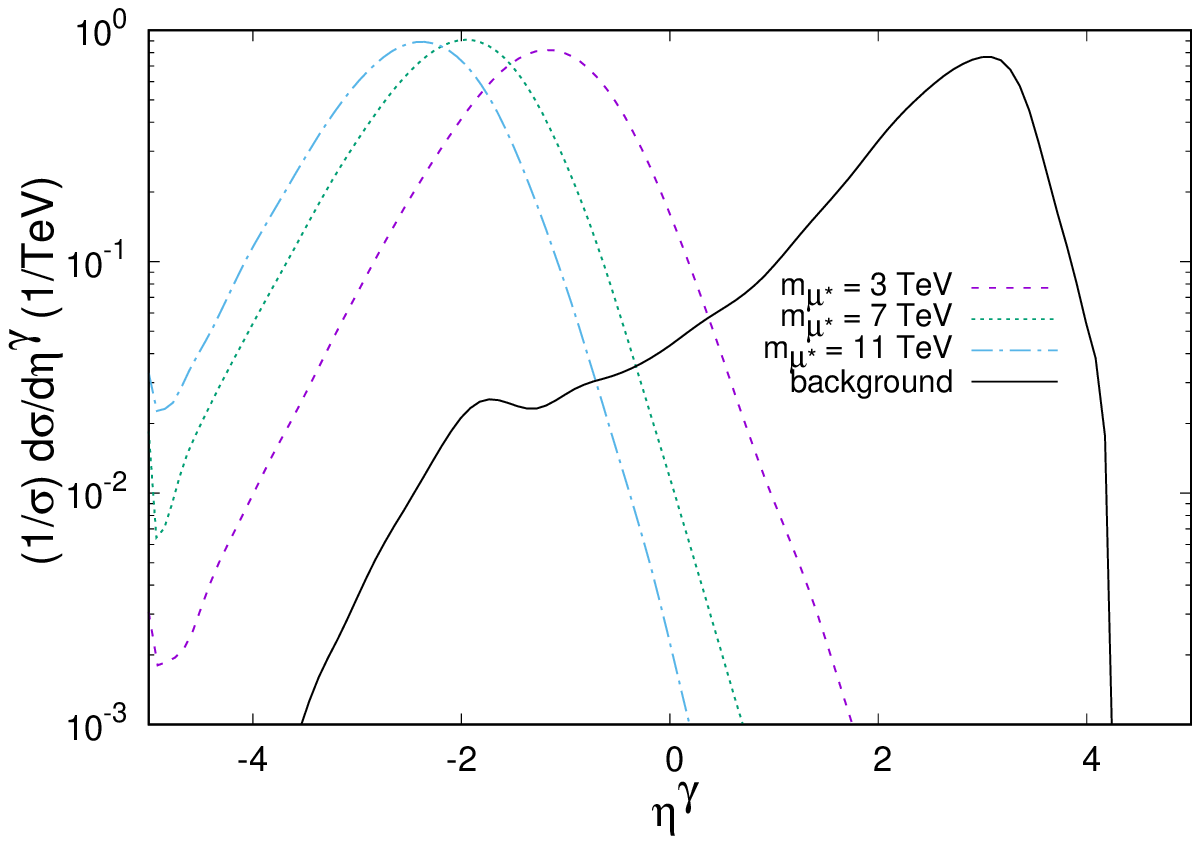}

\caption{The normalized pseudorapidity (left) and transverse momentum (middle)
distributions of the final state muons and the normalized pseudorapidity
distributions of the final state photons (right) at the $\mu750$-SPPC2
collider, for $f=f^{\prime}=1$ and $\varLambda=m_{\mu^{\star}}$.}

\end{figure}

\begin{figure}
\includegraphics[scale=0.45]{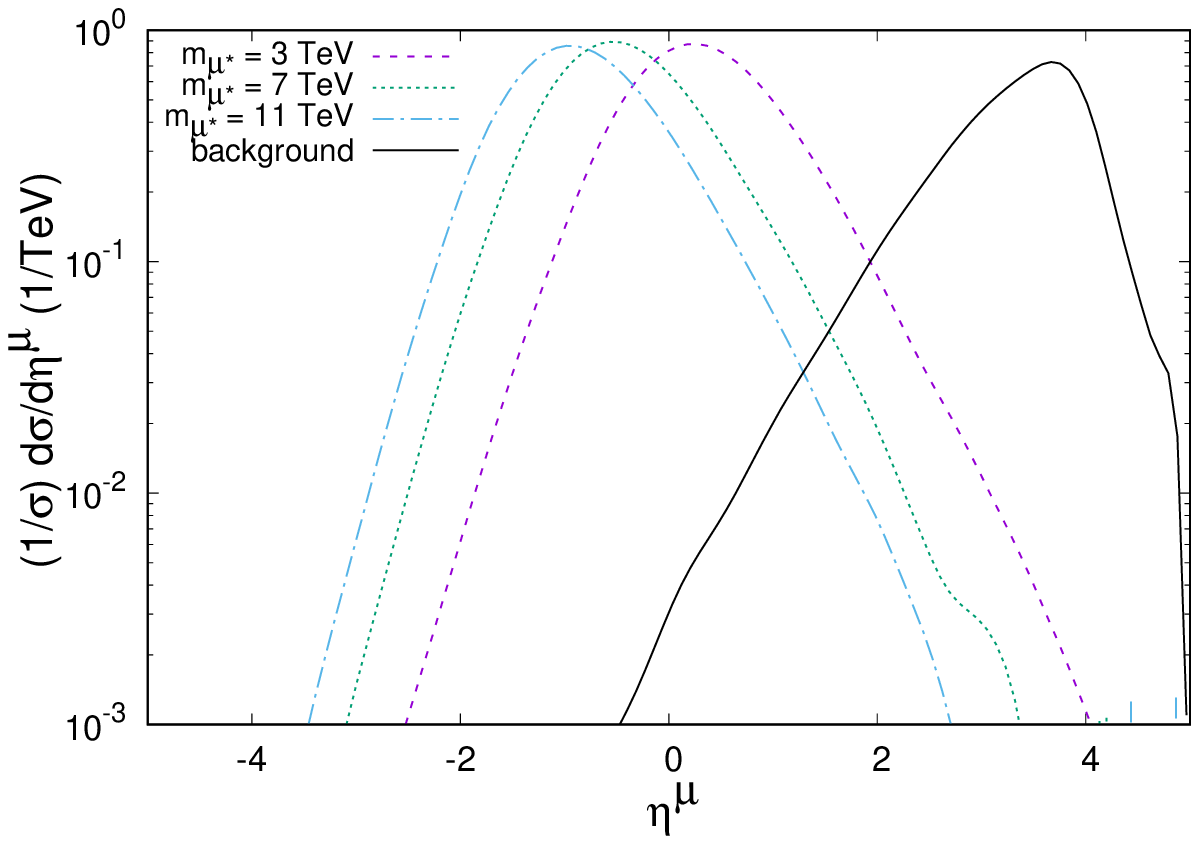}\includegraphics[scale=0.45]{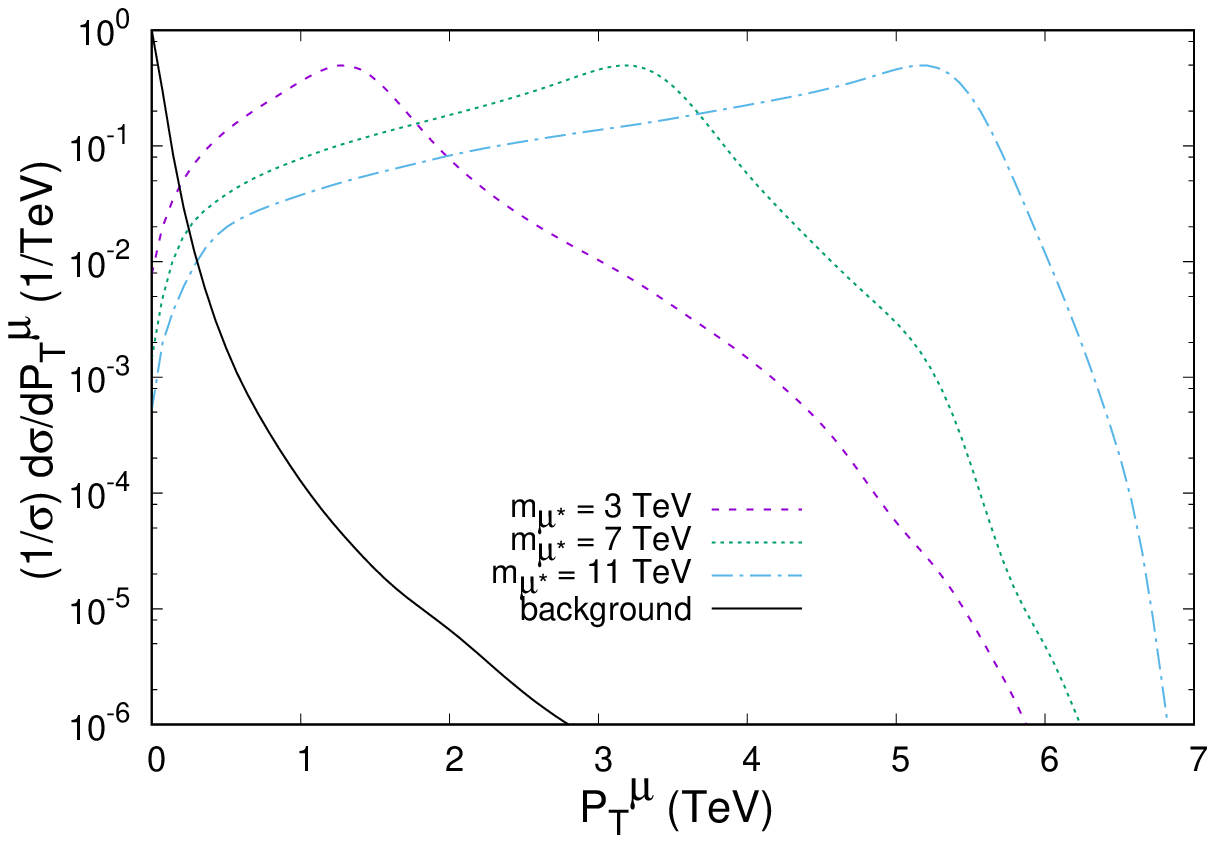}\includegraphics[scale=0.45]{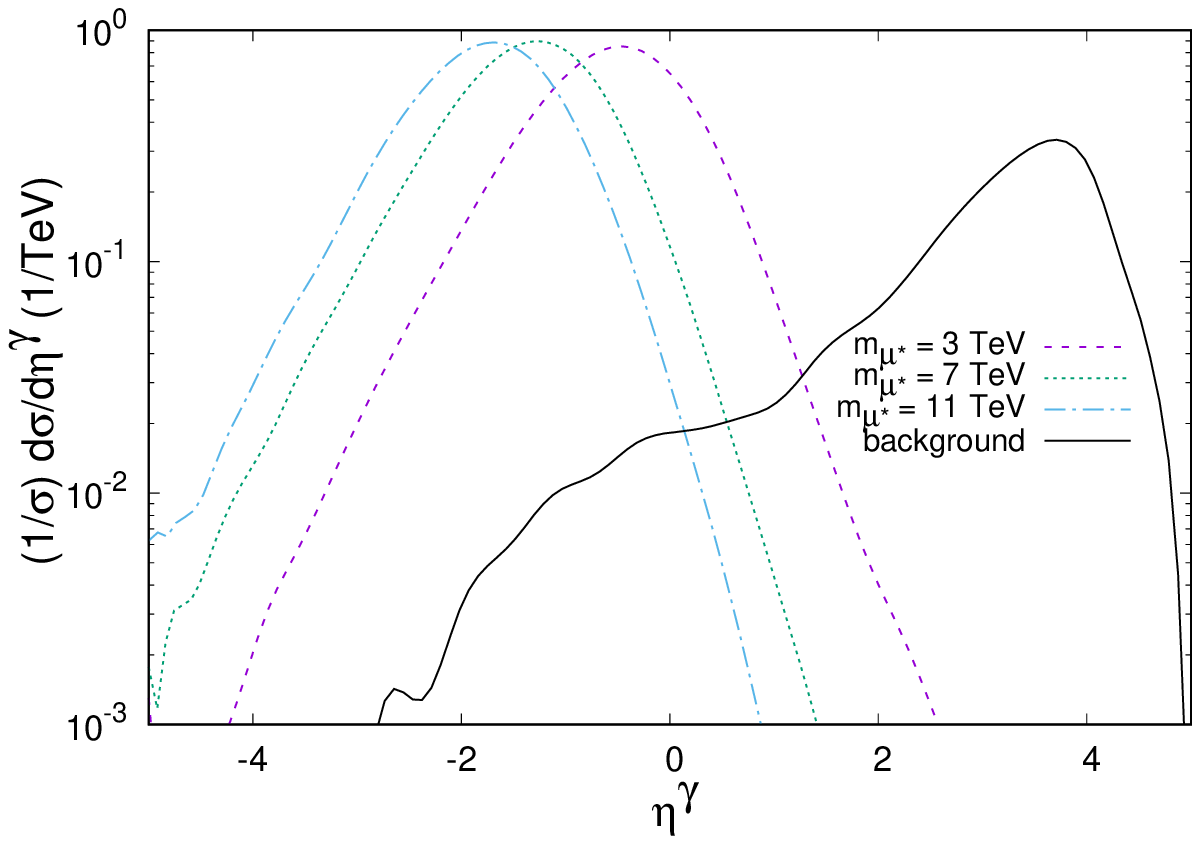}

\caption{The normalized pseudorapidity (left) and transverse momentum (middle)
distributions of the final state muons and the normalized pseudorapidity
distributions of the final state photons (right) at the $\mu1500$-SPPC1
collider, for $f=f^{\prime}=1$ and $\varLambda=m_{\mu^{\star}}$.}
\end{figure}

\begin{figure}
\includegraphics[scale=0.45]{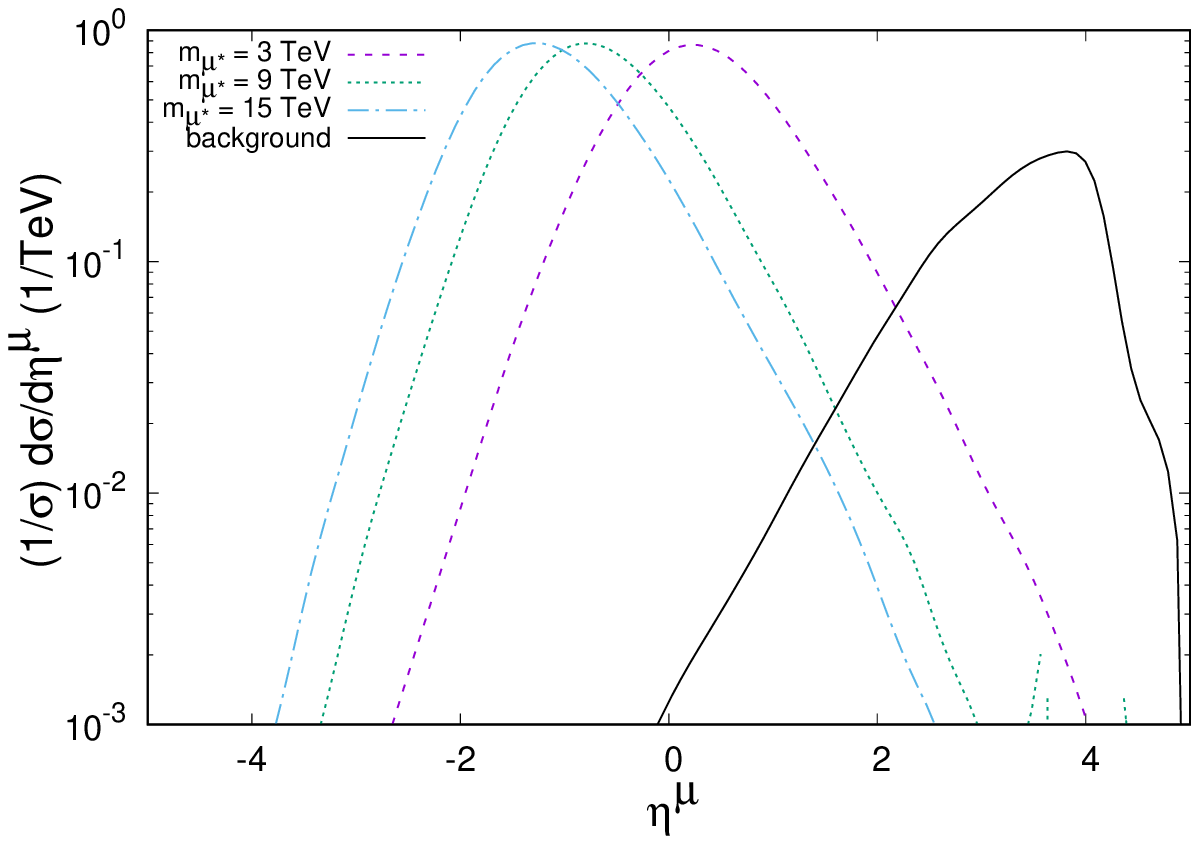}\includegraphics[scale=0.45]{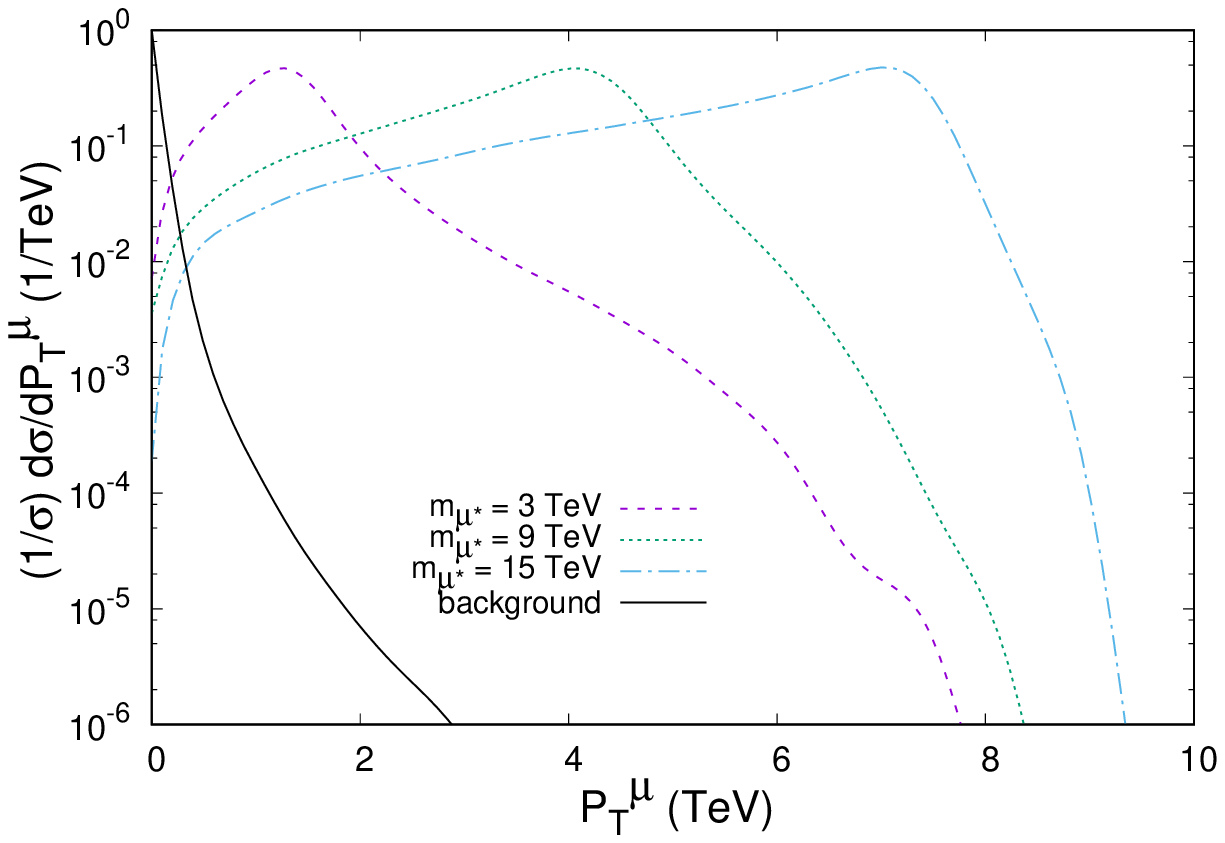}\includegraphics[scale=0.45]{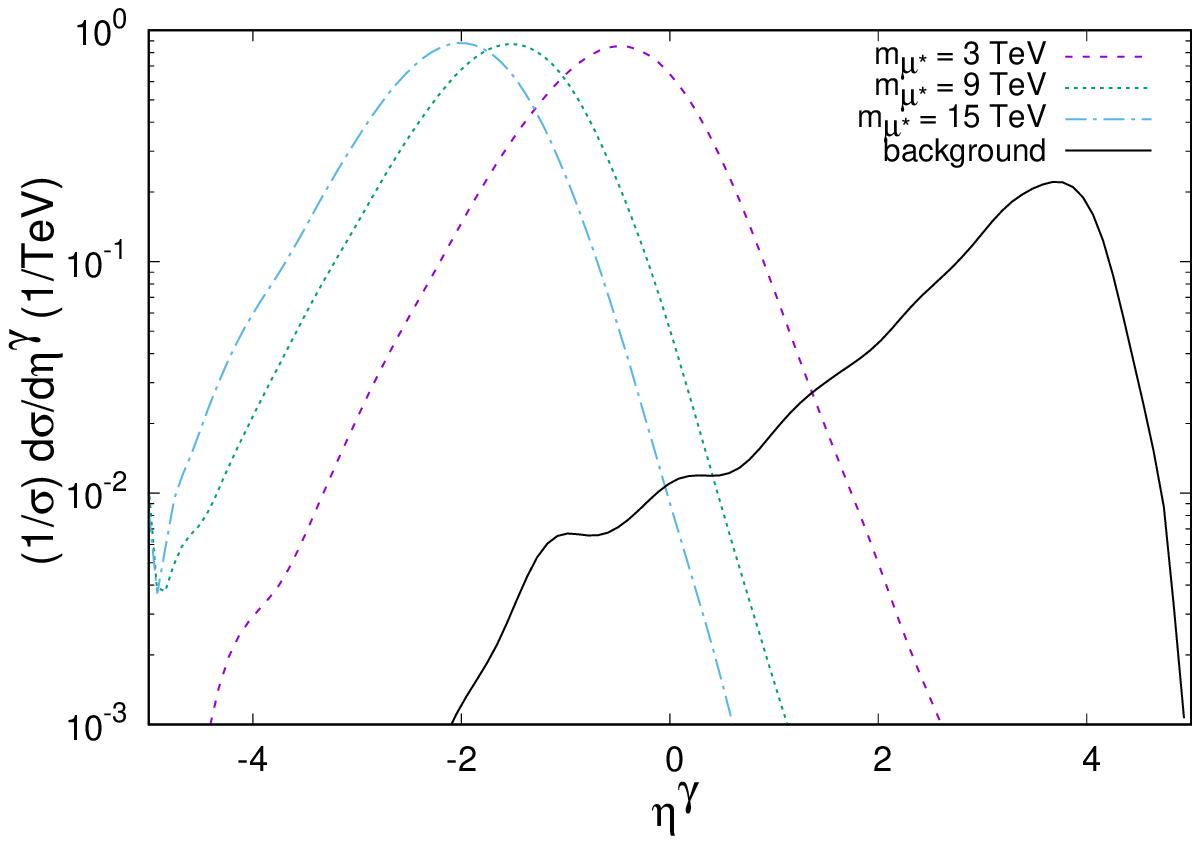}

\caption{The normalized pseudorapidity (left) and transverse momentum (middle)
distributions of the final state muons and the normalized pseudorapidity
distributions of the final state photons (right) at the $\mu1500$-SPPC2
collider, for $f=f^{\prime}=1$ and $\varLambda=m_{\mu^{\star}}$.}

\end{figure}

The normalized $\eta$ and $p_{T}$ distributions of the final state
muons and normalized $\eta$ distributions of the final state photons
are shown in the Fig.3 for $\mu750$-SPPC1, Fig.4 for $\mu750$-SPPC2,
Fig.5 for $\mu1500$-SPPC1 and Fig.6 for $\mu1500$-SPPC2, for both
the signal and background. Since the $p_{T}$ distributions of the
muons are the same as those of the photons for the all colliders,
we have only shown the ones of the muons in these figures. In these
distributions we have chosen optimal regions where we cut off the
most of the background but at the same time keep the signal almost
unchanged. The determined discovery cuts are reported in Table 3.

\begin{table}

\caption{The discovery cuts for the excited muons.}
\begin{tabular}{|c|c|c|c|c|}
\hline 
Colliders & $p_{T}^{\mu}$ & $p_{T}^{\gamma}$ & $\eta^{\mu}$ & $\eta^{\gamma}$\tabularnewline
\hline 
\hline 
 $\mu750$-SPPC1 & $p_{T}^{\mu}$ > $500$ GeV & $p_{T}^{\gamma}$ > $500$ GeV & -5 < $\eta^{\mu}$ < 1 & -4.8 < $\eta^{\gamma}$ < 0\tabularnewline
\hline 
$\mu750$-SPPC2 & $p_{T}^{\mu}$ > $600$ GeV & $p_{T}^{\gamma}$ > $500$ GeV & -5 < $\eta^{\mu}$ < 1.5 & -5 < $\eta^{\gamma}$ < 0.5\tabularnewline
\hline 
 $\mu1500$-SPPC1 & $p_{T}^{\mu}$ > $600$ GeV & $p_{T}^{\gamma}$ > $500$ GeV & -5 < $\eta^{\mu}$ < 2 & -5 < $\eta^{\gamma}$ < 1.5\tabularnewline
\hline 
 $\mu1500$-SPPC2 & $p_{T}^{\mu}$ > $750$ GeV & $p_{T}^{\gamma}$ > $750$ GeV & -5 < $\eta^{\mu}$ < 2.2 & -5 < $\eta^{\gamma}$ < 1.4\tabularnewline
\hline 
\end{tabular}

\end{table}

The invariant mass distributions of the $\mu\gamma$ system after
the application of all discovery cuts are presented in the Fig.7.
As seen clearly from this figure that the separation of the signal
from the background got better.

\begin{figure}
\begin{centering}
\includegraphics[scale=0.5]{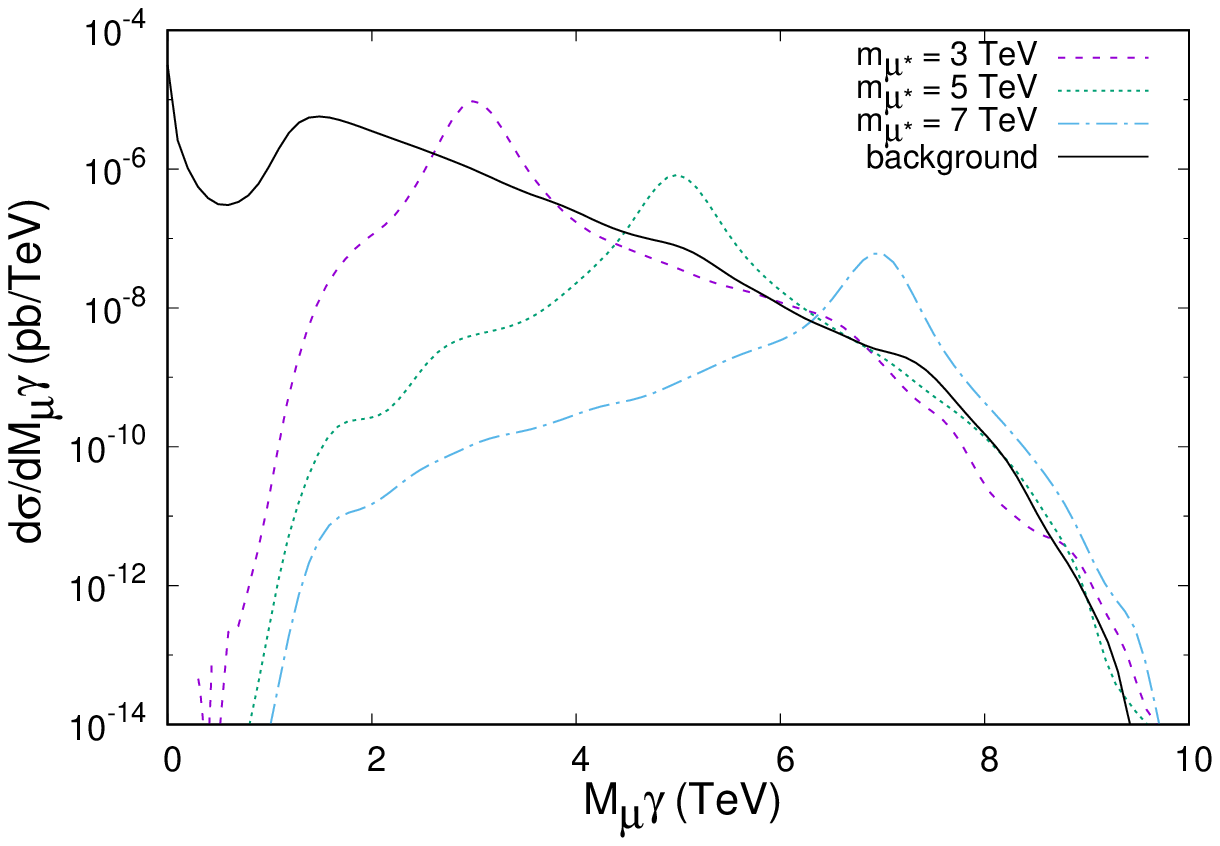}\includegraphics[scale=0.5]{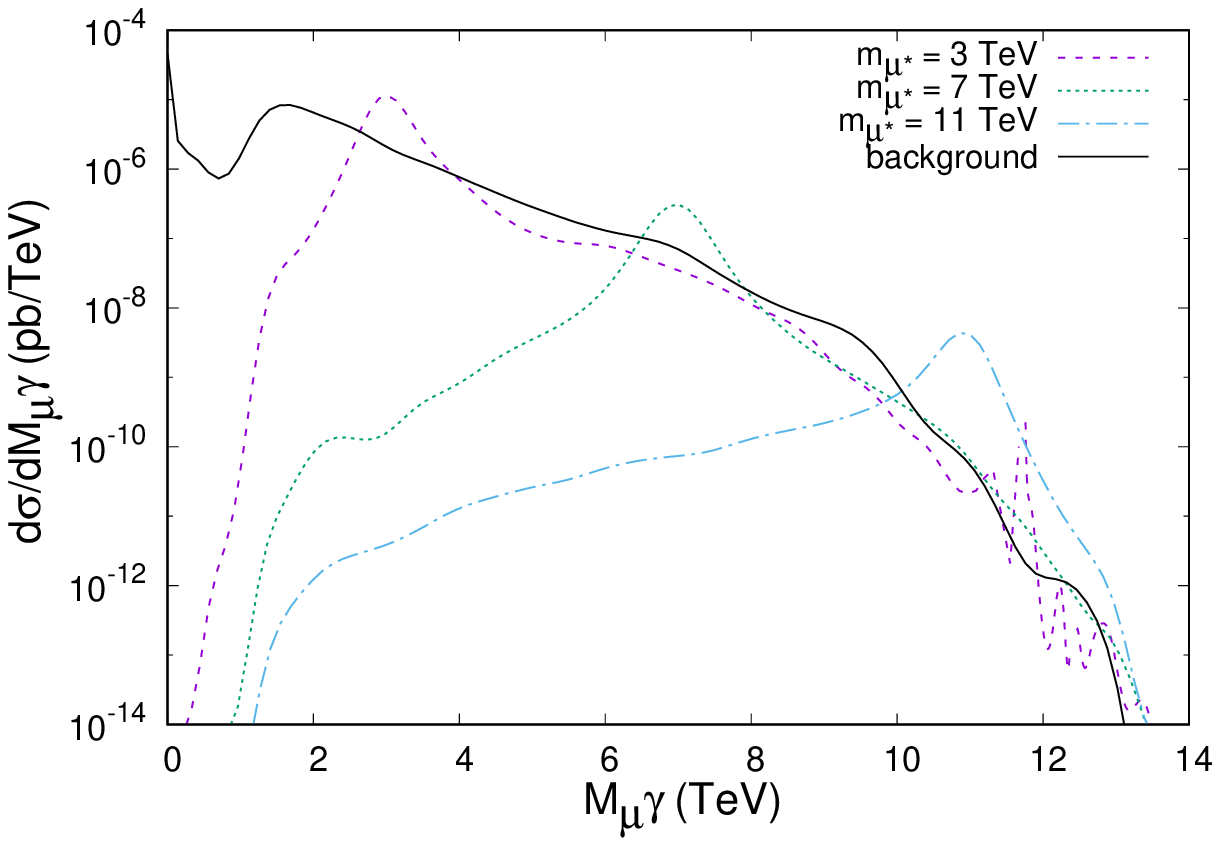}
\par\end{centering}
\begin{centering}
\includegraphics[scale=0.5]{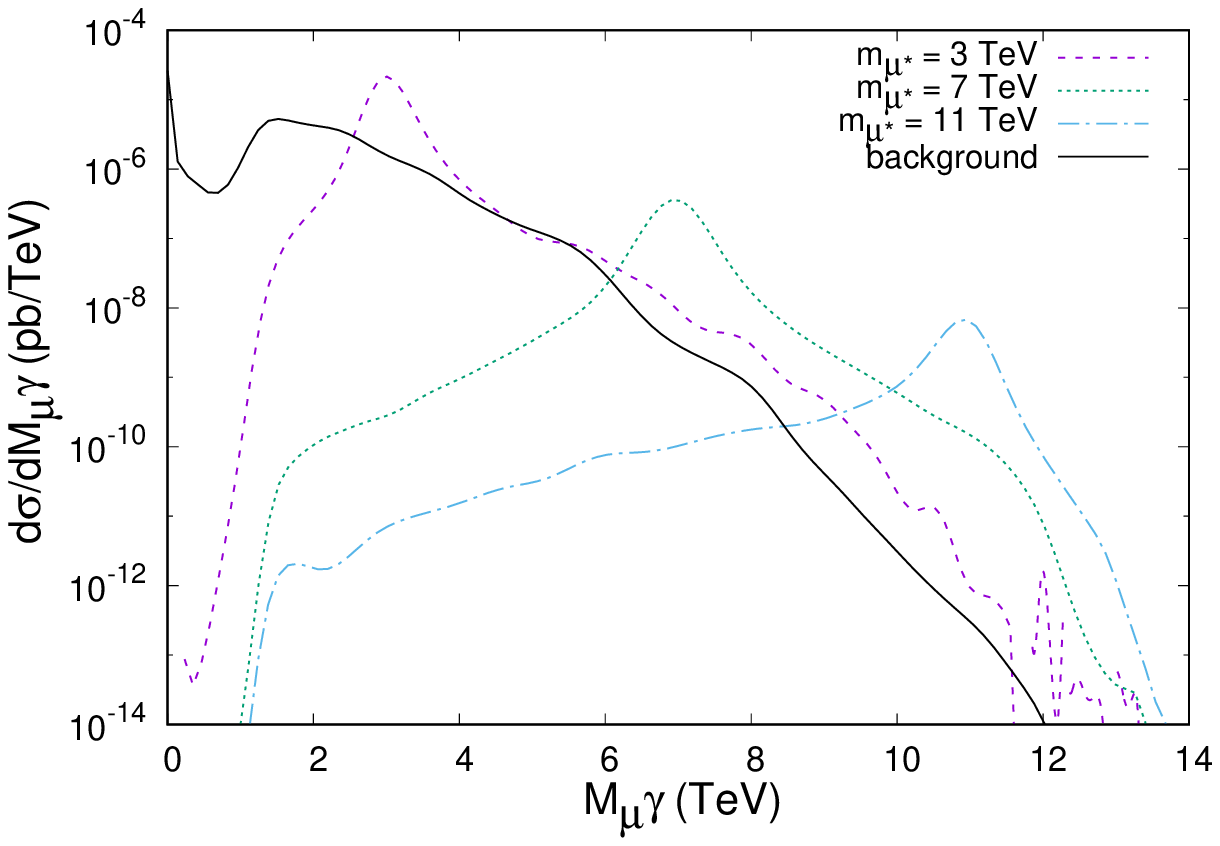}\includegraphics[scale=0.5]{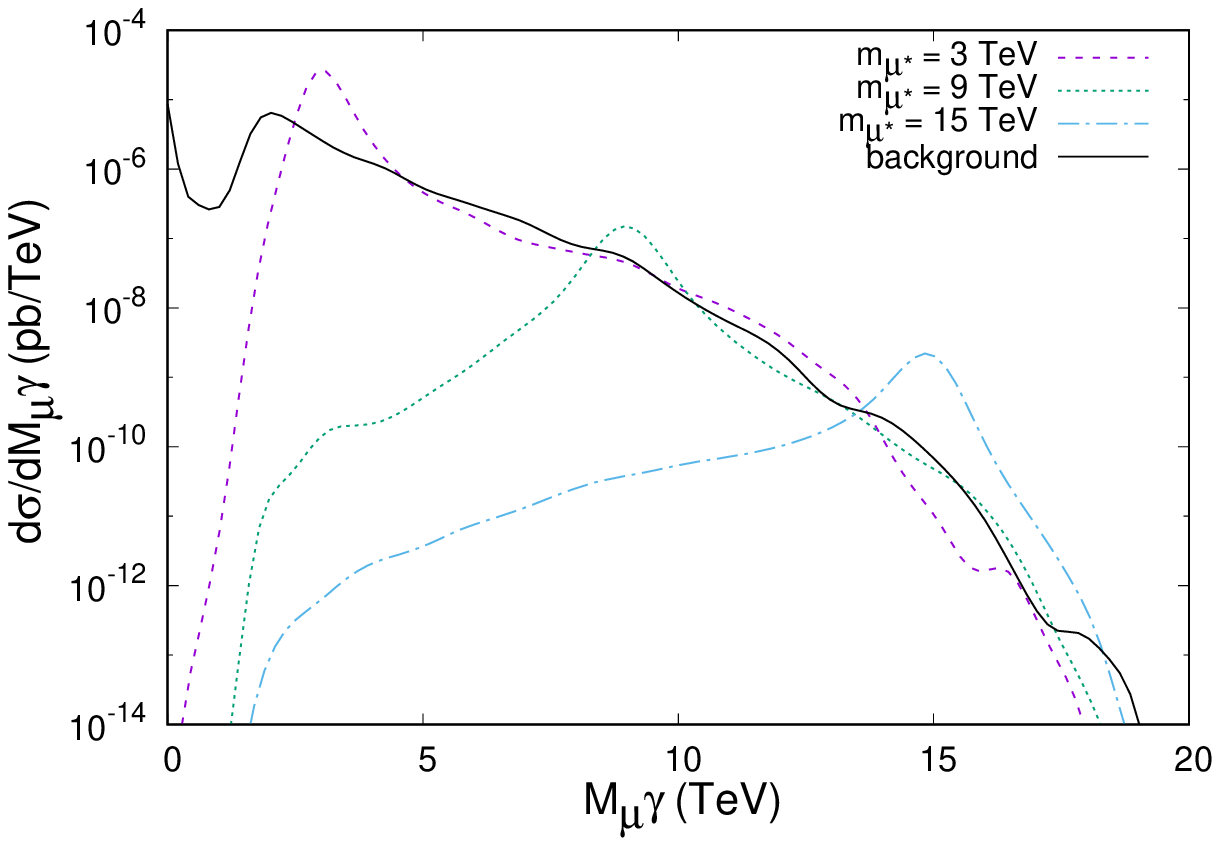}
\par\end{centering}
\caption{The invariant mass distributions of the excited muon signal and the
corresponding background for $\varLambda=m_{\mu^{\star}}$ and $f=f^{\prime}=1$
at the colliders of $\mu750$-SPPC1 (top-left), $\mu750$-SPPC2 (top-right),
$\mu1500$-SPPC1 (bottom-left) and $\mu1500$-SPPC2 (bottom-right).}

\end{figure}

To extract the excited muon signal from the background, in addition
to the discovery cuts, we have imposed a cut on the $\mu\gamma$ invariant
mass as $m_{\mu^{\star}}-2\Gamma_{\mu^{\star}}<m_{\mu\gamma}<m_{\mu^{\star}}+2\Gamma_{\mu^{\star}}$,
where $\Gamma$ denotes the decay width of the excited muon.

For statistical significance (SS) of the excited muon signal, we have
used the following formula,

\begin{equation}
SS=\frac{|\sigma_{S+B}-\sigma_{B}|}{\sqrt{\sigma_{B}}}\sqrt{L_{int}},
\end{equation}

where $\sigma_{S+B}$ denotes the cross section from the signal and
the background, $\sigma_{B}$ denotes the background cross section,
and $L_{int}$ is the integrated luminosity of the collider. Taking
into account the discovery criterion $SS\geq5$, discovery mass limits
on the excited muon have been calculated as 6600, 8700, 7200, 11100
GeV for the colliders of $\mu750$-SPPC1, $\mu750$-SPPC2, $\mu1500$-SPPC1,
$\mu1500$-SPPC2, respectively, assuming $f=f^{\prime}=1$ and $\varLambda=m_{\mu^{\star}}$.
For the $\varLambda=100$ TeV and the same criteria, excited muon
mass limits are 2780, 3900, 3100 and 6700 GeV, respectively also.

\section{CONCLUSION}

We have performed a search for production of the excited muon at the
SPPC-based muon-proton colliders. This work has shown that the these
colliders have a great research potential for the excited muon searches.
We give a realistic estimate for the excited muon signal and the corresponding
background at the SPPC-based muon-proton colliders, namely the $\mu750$-SPPC1
($\sqrt{s}=10.33$ TeV), the $\mu750$-SPPC2 ($\sqrt{s}=14.28$ TeV),
the $\mu1500$-SPPC1 ($\sqrt{s}=14.61$ TeV) and the $\mu1500$-SPPC2
($\sqrt{s}=20.2$ TeV). In the simulations made to obtain the pseudorapidity
and transverse momentum distributions, it is assumed that the energy
scale is $\varLambda=m_{\mu^{\star}}$ and the coupling parameter
is $f=f^{\prime}=1$. The mass limits for exclusion, observation,
and discovery of the excited muons at the four colliders are reported
in Table IV, for both $\varLambda=m_{\mu^{\star}}$ and $\varLambda=100$
TeV. As a result, the future SPPC-based muon-proton colliders offer
the possibility to investigate the excited muon in a very wide range
of mass.

\begin{table}[H]
\caption{The mass limits for the exclusion ($2\sigma$), observation ($3\sigma$),
and discovery ($5\sigma$) of the excited muons at the SPPC-based
$\mu p$ colliders assuming the coupling $f=f^{\prime}=1$ .}
\centering{}%
\begin{tabular}{|c|c|c|c|c|c|}
\hline 
\multirow{1}{*}{Colliders} & \multirow{1}{*}{$L_{int}$($fb^{-1}$) } & $\Lambda$ & $2\sigma$ (GeV) & $3\sigma$ (GeV) & $5\sigma$ (GeV)\tabularnewline
\hline 
\multirow{2}{*}{ $\mu750$-SPPC1} & \multirow{2}{*}{5.5} & $m_{\mu^{\star}}$ & 7400 & 7000 & 6600\tabularnewline
\cline{3-6} 
 &  & $100$ TeV & 4010 & 3750 & 2780\tabularnewline
\hline 
\multirow{2}{*}{$\mu750$-SPPC2} & \multirow{2}{*}{12.5} & $m_{\mu^{\star}}$ & 10600 & 9200 & 8700\tabularnewline
\cline{3-6} 
 &  & $100$ TeV & 6500 & 5000 & 3900\tabularnewline
\hline 
\multirow{2}{*}{ $\mu1500$-SPPC1} & \multirow{2}{*}{4.9} & $m_{\mu^{\star}}$ & 9000 & 8600 & 7200\tabularnewline
\cline{3-6} 
 &  & $100$ TeV & 5000 & 4400 & 3100\tabularnewline
\hline 
\multirow{2}{*}{ $\mu1500$-SPPC2} & \multirow{2}{*}{42.8} & $m_{\mu^{\star}}$ & 13200 & 12600 & 11100\tabularnewline
\cline{3-6} 
 &  & $100$ TeV & 8700 & 8100 & 6700\tabularnewline
\hline 
\end{tabular}
\end{table}
\begin{acknowledgments}
I would like to thank A. Ozansoy for model file support. This work
has been supported by the Scientific and Technological Research Council
of Turkey (TUBITAK) under the grant no 114F337.
\end{acknowledgments}

\end{document}